\documentclass[11pt,a4paper]{JHEP3}

\usepackage{epsf,epsfig,psfrag}
\def\be{\begin{equation}}
\def\ee{\end{equation}}
\newcommand{\bea}{\begin{eqnarray}}
\newcommand{\eea}{\end{eqnarray}}
\newcommand{\nn}{\nonumber}

\newcommand{\bra}[1]{\left\langle #1 \right|}
\newcommand{\ket}[1]{\left| #1 \right\rangle}

\vspace{\baselineskip}

\preprint{\vbox{\hbox{BARI-TH/2010-634 \hfill}
               }}
                
\title{Nonleptonic $B_s$ to charmonium decays:  analyses in pursuit of  determining  the weak phase $\beta_s$}

\author{Pietro Colangelo\\ Istituto Nazionale di Fisica Nucleare, Sezione di Bari, Italy\\ E-mail: \email{pietro.colangelo@ba.infn.it}}
\author{Fulvia De Fazio\\ Istituto Nazionale di Fisica Nucleare, Sezione di Bari, Italy\\ E-mail: \email{fulvia.defazio@ba.infn.it}}
\author{Wei Wang\\ Istituto Nazionale di Fisica Nucleare, Sezione di Bari, Italy\\ E-mail: \email{wei.wang@ba.infn.it}}
\abstract{We analyze nonleptonic $B_s$ decays to a charmonium state and a light  meson,
 induced by the $b \to c {\bar c}s$ transition,  which are  useful to access the $B_s$-${\bar B}_s$ mixing phase $\beta_s$. We use generalized factorization and  $SU(3)_F$ symmetry to relate such modes to correspondent $B$ decay channels. We discuss the feasibility of the measurements in the various channels, stressing the importance of comparing  different determinations of $\beta_s$ in view of the hints of new physics effects (NP) recently emerged in the $B_s$ sector. Finally, adopting a general parametrization of  NP  contributions  to the decay amplitudes, we discuss how to experimentally constrain  new physics parameters.}
\begin{document}

\clearpage
\setcounter{page}{1}

\section{Introduction}\label{sec:intro}
The detailed analysis of CP violation in particle physics is a powerful tool to test the standard model (SM) of elementary interactions and enveil the effects of new interactions.
The fundamental role in the SM description of CP violation is played by the Cabibbo-Kobayashi-Maskawa (CKM) mixing matrix, which is unitary and implies CP violation  if it  is
complex.  The constraints stemming from  unitarity can be represented as triangles,  the lengths of whose
sides are the moduli of products of CKM elements,  while the angles
represent relative phases between them.  The most studied $bd$ unitarity triangle is defined by the relation
$V_{ud}V_{ub}^*+V_{cd}V_{cb}^*+V_{td}V_{tb}^*=0$,  and   has been probed mainly through the extensive analysis of $B_d$
phenomenology.  As a result,   the  CKM parameters  in the Wolfenstein parameterization have
been fixed with  small errors  through the  measurement of the sides and the angles of this  triangle \cite{fits}.  The next, already ongoing, effort  is to look at processes in which to
test the SM  requires a greater experimental and theoretical precision. The $B_s$ sector  is suitable for such a purpose.  As in the $B_d$ case, one of the CKM
unitarity constraints involves  matrix elements related to $B_s$
decays: $V_{us}V_{ub}^*+V_{cs}V_{cb}^*+V_{ts}V_{tb}^*=0$,   and one of the angles of the corresponding    $bs$
triangle is  the phase of the $B_s$-${\bar B}_s$ mixing:
$\beta_s=Arg\left[-{V_{ts}V_{tb}^* \over V_{cs} V_{cb}^*}\right]$.  In  SM  $\beta_s$  is
expected to be tiny:  $\beta_s \simeq 0.017$ rad.

$B_s$ is produced at the $B$ factories running at the peak
of $\Upsilon(5S)$ and in hadron collisions. In particular, the experiments CDF and D0 at the Tevatron have
obtained a number of remarkable  results, such as  the  measurement of the mixing parameters: The mass difference of the two $B_s$ mass eigenstates has been fixed to $\Delta m_s=17.77 \pm 0.10 (\rm stat) \pm 0.07 (\rm syst)$ ps$^{-1}$ \cite{Abulencia:2006ze} , while the value of their  width difference $\Delta \Gamma_s$ depends on the constraints  on $\beta_s$ adopted in the experimental analysis \cite{Abazov:2007tx};  noticeably, $\Delta \Gamma_s$
   is not small as in the $B_d$ case.
Furthermore,  these Collaborations have provided us with results which seem to signal  new physics (NP) effects.
The first one
 concerns the phase $\beta_s$, extracted from the angular
analysis of the time-dependent differential decay width in the
process $B_s \to J/\psi \phi$. The study is rather involved: an
angular analysis is needed to disentangle the CP-even and CP-odd
components, required  since the final state of two vector mesons is
not a CP eigenstate.  Moreover, the measurement can be carried out either
considering flavour tagged  or  untagged decays. Another issue  concerns the use or not of assumptions on the
strong phases among the different helicity amplitudes in the
considered process: this assumption has been once adopted by D0
Collaboration  in one study \cite{:2008fj}. Different results have been
obtained from the different analyses  \cite{:2008fj,Abazov:2007zj}, and,
averaging them, the Heavy Flavour Averaging Group has provided  a value of $\beta_s$ consistent with SM only at $2.2$ $\sigma$ level:
$\phi_s^{J/\psi \phi}=-2 \beta_s=-0.77 \pm^{0.29}_{0.37}$ or
$\phi_s^{J/\psi \phi}=-2 \beta_s=-2.36 \pm^{0.37}_{0.29}$
\cite{Barberio:2008fa}. A   new measurement announced by CDF: $\beta_s \in [0.0,0.5] \cup \,[1.1,1.5]$  (at 68$\%$ CL) \cite{talkcdf}, if confirmed, would  reconcile the SM prediction with experiment.

Another signal of a  possible inadequacy of the SM  is the measurement of an anomalous like-sign dimuon charge asymmetry of semileptonic $b$-hadron decay,  reported by the D0 Collaboration \cite{Abazov:2010hv} (updating
a previous measurement  \cite{Abazov:2006qw}):
\be
A_{sl}^b=\frac{N_b^{++}-N_b^{--}}{N_b^{++}+N_b^{--}} = -(9.57 \pm 2.51 \pm 1.46) \times 10^{-3} \label{asym-dimuon} \,.\ee
Hence, there is a large excess of negatively charged muons over positively charged ones which would have been generated by the oscillation of one neutral $b$ meson into the other, at odds  with the SM expectation
$A_{sl}^b=\left( -0.310^{+0.083}_{-0.098} \right)\times 10^{-3}$ \cite{Lenz:2010gu}, a result which might imply a NP  effect  in the oscillation.

In this complex scenario  it is worth to further analyze   the $B_s$ sector, trying to identify and reduce the  uncertainties affecting the theoretical predictions, with the aim of
improving the measurement of $\beta_s$,    overcoming  the
difficulties in $B_s \to J/\psi \phi$. Notice that this channel is  considered a golden mode,  since it is  induced by the $b \to c {\bar c}s$ transition in which, in SM,    the only weak phase involved is
that of the mixing, so that  the indirect CP asymmetry would be proportional to $\sin(2 \beta_s)$, much in the same way as $B_d \to
J/\psi K_S$ has provided a  determination of the angle $\beta$. The feasibility in the reconstruction of the products
of the subsequent decays $J/\psi \to \mu^+ \mu^-$, $\phi \to K^+ K^-$ makes this channel also experimentally appealing.

There are  other modes that can be used to access $\beta_s$, namely $B_s \to M_{(c{\bar c})} + L$, where $M_{(c{\bar c})}$ is a charmonium state  $J/\psi$, $\Psi(2S)$, $\eta_c$,   $\eta_c(2S)$, $\chi_{c0}$, $\chi_{c1}$, $\chi_{c2}$, $h_c$ and $L$ is a light scalar, pseudoscalar or vector meson, $f_0(980)$, $\eta$, $\eta^\prime$ and $\phi$. Each of these channels
presents specific features and advantages/difficulties which we want to discuss here. Standing the general theoretical difficulty
in  the calculation of nonleptonic decay amplitudes,  in the next
Section we  discuss  approaches   to afford the problem,  and   exploit the generalized factorization  to calculate the  branching fractions in the SM. In this way  the  most suitable processes to determine $\beta_s$ can be identified.
 In Section \ref{sect:np} we also consider the possible impact of new physics in  these modes, and discuss how to exploit experimental data  to constrain  NP parameters.
 Conclusions are presented  in the last Section.

\section{$B_s \to M_{c{\bar c}} L$ decays}
\label{decays}
In SM,  the effective hamiltonian governing the
nonleptonic decays induced by the $b \to c {\bar c}s$ transition reads as \cite{Buchalla:1995vs}:
 \begin{eqnarray}
 {\cal H}_{\rm eff} &=& \frac{G_{F}}{\sqrt{2}}
     \bigg\{  V_{cb} V_{cs}^{*} \big[
     C_{1}({\mu}) O_{1}(\mu)
  +  C_{2}({\mu}) O_{2}(\mu)\Big] -V_{tb} V_{ts}^{*} \Big[{\sum\limits_{i=3}^{10,7\gamma,8g}} C_{i}({\mu}) O_{i}({\mu})
  \big ] \bigg\} ,
 \label{eq:hamiltonian-b-ccs}
\end{eqnarray}
where $G_F$ is the Fermi constant, the $C_i$ are Wilson coefficients,  and  $O_i$ are
\begin{itemize}
\item current--current (tree) operators
\begin{eqnarray}
O_1= {\bar c} \gamma_\mu (1-\gamma_5)b \,\,  {\bar s} \gamma^\mu (1-\gamma_5)c,
 \ \ \ \ \ \ \ \ \
 O_2= {\bar c} \gamma_\mu (1-\gamma_5)c \,\,  {\bar s} \gamma^\mu (1-\gamma_5)b
 \end{eqnarray}
     \item  QCD penguin operators
    \begin{eqnarray}
      O_{3}=({\overline{s}}_{\alpha}b_{\alpha})_{V-A}\sum\limits_{q^{\prime}}
           ({\overline{q}}^{\prime}_{\beta} q^{\prime}_{\beta} )_{V-A},
    \ \ \ \ \ \ \ \ \
    O_{4}=({\overline{s}}_{\beta} b_{\alpha})_{V-A}\sum\limits_{q^{\prime}}
           ({\overline{q}}^{\prime}_{\alpha}q^{\prime}_{\beta} )_{V-A}, \nonumber \\
     \!\!\!\! \!\!\!\! \!\!\!\! \!\!\!\! \!\!\!\! \!\!\!\!
    O_{5}=({\overline{s}}_{\alpha}b_{\alpha})_{V-A}\sum\limits_{q^{\prime}}
           ({\overline{q}}^{\prime}_{\beta} q^{\prime}_{\beta} )_{V+A},
    \ \ \ \ \ \ \ \ \
    O_{6}=({\overline{s}}_{\beta} b_{\alpha})_{V-A}\sum\limits_{q^{\prime}}
           ({\overline{q}}^{\prime}_{\alpha}q^{\prime}_{\beta} )_{V+A},
    \label{eq:operator56}
    \end{eqnarray}
 \item electro-weak penguin operators
    \begin{eqnarray}
     O_{7}=\frac{3}{2}({\overline{s}}_{\alpha}b_{\alpha})_{V-A}
           \sum\limits_{q^{\prime}}e_{q^{\prime}}
           ({\overline{q}}^{\prime}_{\beta} q^{\prime}_{\beta} )_{V+A},
    \ \ \ \
    O_{8}=\frac{3}{2}({\overline{s}}_{\beta} b_{\alpha})_{V-A}
           \sum\limits_{q^{\prime}}e_{q^{\prime}}
           ({\overline{q}}^{\prime}_{\alpha}q^{\prime}_{\beta} )_{V+A}, \nonumber \\
     O_{9}=\frac{3}{2}({\overline{s}}_{\alpha}b_{\alpha})_{V-A}
           \sum\limits_{q^{\prime}}e_{q^{\prime}}
           ({\overline{q}}^{\prime}_{\beta} q^{\prime}_{\beta} )_{V-A},
    \ \ \ \
    O_{10}=\frac{3}{2}({\overline{s}}_{\beta} b_{\alpha})_{V-A}
           \sum\limits_{q^{\prime}}e_{q^{\prime}}
           ({\overline{q}}^{\prime}_{\alpha}q^{\prime}_{\beta} )_{V-A},
    \label{eq:operator9x}
    \end{eqnarray}
     \item magnetic moment operators
    \begin{eqnarray}
     O_{7\gamma}&=&-\frac{e}{4\pi^2}{\overline{s}}_{\alpha}\sigma^{\mu\nu}(m_s P_L+m_b
     P_R)b_{\alpha}F_{\mu\nu}, \nonumber \\
    O_{8g}&=&-\frac{g}{4\pi^2}{\overline{s}}_{\alpha}\sigma^{\mu\nu}(m_s P_L+m_b
     P_R)T^a_{\alpha\beta}b_{\beta}G^a_{\mu\nu}.\label{eq:operator7gamma8g}
\end{eqnarray}
\end{itemize}
$\alpha$ and $\beta$ are color indices,  and $q^\prime$ are the
active $q^\prime=(u,d,s,c,b)$ quark fields at the scale $m_b$ with charge $e_{q^\prime}$. The
right (left) handed current is defined as $({\overline{q}}^{\prime}_{\alpha}
q^{\prime}_{\beta} )_{V\pm A}= {\overline{q}}^{\prime}_{\alpha}
\gamma_\nu (1\pm \gamma_5) q^{\prime}_{\beta}  $,  with projection operators
$P_{R,L}=\frac{1\pm\gamma_5}{2}$.
Assuming  CKM unitarity and neglecting  the tiny product
$V_{ub}V_{us}^*$,  the relation holds:  $V_{tb} V_{ts}^{*}=-V_{cb} V_{cs}^{*}$.

The  hamiltonian  (\ref{eq:hamiltonian-b-ccs}) induces the decays of  $SU(3)_F$ related states,
namely the decays of  $B_d$,  $B^-$ and $B_s$;
 we  consider  the general case of the decay of  a $B_a$ meson ($a=u,d,s$ being the light flavour index).
The simplest approach to compute the matrix element of the  hamiltonian (\ref{eq:hamiltonian-b-ccs}) between given initial and final hadronic states is the naive factorization approach. In such an approach, neglecting the magnetic moment operators in (\ref{eq:operator7gamma8g}), the   $B_a \to M_{c {\bar c}} L$  amplitude  reads:
\be
{\cal A}({\bar B}_a \to M_{c {\bar c}} L)={G_F \over \sqrt{2}} V_{cb}V_{cs}^* a_2^{eff}(\mu)
\bra{M_{(c{\bar c})}} {\bar
c} \gamma^\mu(1-\gamma_5) c \, \ket{0}\bra{L} {\bar s}
\gamma_\mu(1-\gamma_5) b \ket{{\bar B}_a} \,, \label{amplitude-compl} \ee
where $a_2^{eff}(\mu)=a_2(\mu)+a_3(\mu)+a_5(\mu)$ and $a_2=C_2+\displaystyle{C_1 \over N_c}$, $a_3=C_3+\displaystyle{C_4 \over N_c}+\displaystyle{3 \over 2}e_c \left(C_9+\displaystyle{C_{10} \over N_c} \right)$ and $a_5=C_5+\displaystyle{C_6 \over N_c}+\displaystyle{3 \over 2}e_c \left(C_7+\displaystyle{C_{8} \over N_c} \right)$.
However, naive factorization predictions are not able to reproduce several branching ratios for which experimental data are available. Among these there are $B_d$ decays induced by the transition $b \to c {\bar c}s$:  some of these modes, which are of interest for the  present analysis, are listed in Table~\ref{tab:experimental-data} together with the  experimental branching fractions.
%%%%%%%%%%%%%%%%%%%%%%%%%%%%%%%%%%%%%%%%%%%%%%%%%%%%%%%%%%%%%%%%%%%%%%%%%%%%%%%%%%%%%%%%%%%%%%%%%
\TABULAR{cccccc}
 {\hline
  $B\to M_{\bar cc} K$ & $J/\psi$ & $\eta_c$   & $\psi(2S)$ & $\eta_c(2S)$ \\
 $B^-$ & $10.07\pm0.35$   & $9.1\pm1.3$ &   $6.48\pm0.35$ &  $3.4\pm1.8$   \\
 $B^0$ & $8.71\pm0.32$ & $8.9\pm1.6$  &  $6.2\pm0.6$ &  \\
 \hline
  $B\to M_{\bar cc} K^*$ & $J/\psi$ & $\eta_c$  & $\psi(2S)$ & $\eta_c(2S)$ \\
 $B^-$ & $14.3\pm0.8$   & $12.0\pm7.0$  & $6.7\pm1.4$ &    \\
 $B^0$ & $13.3\pm0.6$ & $9.6\pm3.3$   &  $7.2\pm0.8$ & $<3.9$\\
 \hline
  $B\to M_{\bar cc}  K$ &    $\chi_{c0}$ & $\chi_{c1}$ & $\chi_{c2}$ &
 $h_c$ \\
 $B^-$   & $1.43\pm0.21$ &
 $5.1\pm0.5$ & $<0.29$ & $<0.38$  \\
 $B^0$   & $<1.13$ & $3.9\pm0.4$ &
 $<0.26$  & \\
 \hline
  $B\to M_{\bar cc} K^*$   & $\chi_{c0}$ & $\chi_{c1}$ & $\chi_{c2}$ &
 $h_c$  \\
 $B^-$   & $<2.1$ &
 $3.6\pm0.9$ & $<0.12$ &  \\
 $B^0$   & $1.70\pm0.40$ & $2.0\pm0.6$ &
 $<0.36(0.66\pm0.19)$  & $(<2.2)$ \\
 \hline}
{Experimental  results    for  the branching fractions ${\cal B}(B\to M_{\bar cc} K^{(*)})$   ($\times 10^{4}$)  \cite{Amsler:2008zzb}; results in parentheses are from \cite{Barberio:2008fa}.\label{tab:experimental-data}}
%%%%%%%%%%%%%%%%%%%%%%%%%%%%%%%%%%%%%%%%%%%%%%%%%%%%%%%%%%%%%%%%%%%%%%%%%%%%%%%%%%%%%%%%%%%%%%%%%%%%%%

Several modifications of the  naive factorization ansatz have been
proposed. One possibility is to consider the Wilson
coefficients as effective parameters to be determined from experiment \cite{Neubert:1997uc}. In principle,  this implies that  such coefficients are
channel-dependent. However,  some channels could be related, namely
invoking flavour symmetries, so that universal values for the coefficients can be assumed
within a certain class of
modes.  In our case, this {\it generalized} factorization approach
consists in considering the quantity $a_2^{eff}$ in (\ref{amplitude-compl}) as a
process-dependent parameter to be fixed from experiment. In
particular, on the basis of  $SU(3)_F$ symmetry, $B_q$ ($B_u$ or $B_d$)
decays can be related to analogous $B_s$ decays induced by the
same $b \to c {\bar c}s$ transition, so that experimental data concerning
$B_q$ modes provide  predictions for $B_s$ related ones.
Also this method presents some drawbacks, for example  the issue of rescattering in the final state and of the strong phases in the various amplitudes cannot be
faced \cite{Buras:1998us}. Nevertheless, it is
 useful from a phenomenological point of view, at least to understand the size of nonleptonic branching ratios.

A different procedure to analyze nonleptonic decays is the
hard-scattering approach,  based on the assumption of the dominance of hard
gluon exchange and of the suppression of  soft mechanisms  due to low energy gluon exchanges. In this approach
 a  nonleptonic amplitude is  expressed as a
convolution of a  hard kernel, computed in perturbation theory,  with the
light-cone wave functions of the hadrons involved in the decay.  In this so-called
 perturbative QCD approach (pQCD)
the suppression of the soft term is achieved by
 suitable Sudakov suppression factors, but  the uncertainty in
 the wave functions limits the accuracy of the predictions   \cite{Keum:2000ph}.

A systematic improvement of naive factorization is QCD factorization (BBNS) \cite{Beneke:1999br}. In this approach,
a factorization formula is written  for a nonleptonic $B_a \to M_1 M_2$ decay amplitude  ($M_1$ denotes the meson
picking up the $B_a$ spectator quark), valid in the heavy quark limit (i.e. up to
$\Lambda_{QCD}/m_b$ corrections).  This formula reproduces
the naive factorization result at leading order in $\alpha_s$ and $\Lambda_{QCD}/m_b$; however,  it  cannot be applied  when
the meson that does not pick up the $B_a$ spectator quark is heavy.
The knowledge of the meson wave functions is required and represents therefore  a
limiting factor.

A particular case is represented by the decays  $B \to M_{\bar c  c} L$ considered here. Since the charmonium state  is  a heavy meson,
the BBNS factorization formula does not hold. However,  it has been pointed out that, being a charmonium meson
a state with small transverse extension, one can still  adopt
the factorization formula. However  a problem arises  going beyond
the leading twist for the wave functions,  since the factorization formula contains convolution integrals of such wave functions, and higher twist wave
functions do not vanish in the end point,  developing divergences. Getting rid of such divergences
requires the introduction of a cutoff,   a parameter to be fixed from experiment.

Two body nonleptonic $B$ decays have also been analyzed in a modified formulation of light-cone QCD sum rules originally proposed in \cite{Khodjamirian:2000mi}
 to calculate the $B \to \pi \pi$ matrix element, finding results in agreement with QCD factorization.  Applying this approach  to $B$ to charmonium decays, one finds that  nonfactorizable contributions are important, but that their inclusion does not allow to reproduce experimental data for  $B \to J/\psi K$  \cite{Melic:2003bw}.

Hence,  no  satisfactory treatment of nonleptonic $B$ to charmonium decays exists at present, each method having its own advantages/drawbacks.
Standing the  phenomenological importance of these modes, we
afford  a study based on generalized factorization,   aiming at establishing at least the sizes of
 the branching ratios of these modes and  their role for a measurement of  $\beta_s$.

To apply eq. (\ref{amplitude-compl}) to  the modes  we are analyzing, we need  the following hadronic quantities:
\begin{itemize}
\item charmonium decay constants:
\bea
\bra{\eta_c(q)} {\bar c} \gamma_\mu\gamma_5 c \, \ket{0}&=&-i f_{\eta_c} q_\mu \,\,,\nonumber \\
\bra{J/\psi(q,\epsilon)} {\bar c} \gamma_\mu c \, \ket{0}&=&f_\psi m_\psi \epsilon^*_\mu \,\,,\nonumber \\
\bra{\chi_{c1}(q,\epsilon)} {\bar c} \gamma_\mu \gamma_5 c \, \ket{0}&=& f_{\chi_{c1}} m_{\chi_{c1}}\epsilon^*_\mu \,\,,
 \label{decay-constants} \eea
 ($\epsilon(\lambda)$ polarization vector); for $\chi_{c{0,2}}$ and $h_c$ one has
  $\bra{\chi_{c0}(q)} {\bar c} \gamma^\mu c \, \ket{0}=\bra{\chi_{c2}(q,\epsilon)} {\bar c} \gamma^\mu c \, \ket{0}=\bra{h_{c}(q,\epsilon)} {\bar c} \gamma^\mu (1-\gamma_5) c \, \ket{0}= 0$;
\item  ${\bar B}_a \to L$ form factors, with  $L$ a pseudoscalar $(P)$ or a  scalar $(S)$ meson:
\bea
\langle P(S)(p^\prime)|\overline s \gamma^{\mu} (\gamma_5) b| {\bar B}_a(p)\rangle
   &=&F_1(q^2)\left[ (p+p^\prime)^{\mu}-\frac{m_{B_a}^2-m_{P(S)}^2}{q^2}q^{\mu} \right]\nonumber \\
     &+&F_0(q^2)\,\frac{m_{B_a}^2-m_{P(S)}^2}{q^2}q^{\mu} \,\,,\label{B-to-S-P-ff}
     \eea
\item ${\bar B}_a \to L$ form factors,  with $L$ a vector $(V)$ meson:
\bea
 &&\langle L(p^\prime,\epsilon)|\overline s \gamma_{\mu}(1-\gamma_5)b| {\bar B}_a(p)\rangle
={2 V(q^2) \over m_{B_a}+m_L} \epsilon_{\mu \nu \alpha \beta} \epsilon^{* \nu} p^\alpha p^{\prime \beta}  \nonumber \\
&-i & \big[ \epsilon^{* \mu}(m_{B_a}+m_L) A_1(q^2)-(\epsilon^* \cdot q) (p+p^\prime)_{\mu}{A_2(q^2) \over m_{B_a}+m_L} \nonumber \\ && -(\epsilon^* \cdot q) {2 m_L \over q^2} (A_3(q^2)-A_0(q^2))q_\mu \big] \,\,,
\label{B-to-V-ff}
\eea
\end{itemize}
with $\displaystyle{A_3(q^2)={m_{B_a}+m_L \over 2 m_L}A_1(q^2)-{m_{B_a}-m_L \over 2 m_L}A_2(q^2)}$.
The first two equations in (\ref{decay-constants}) also hold for  $\eta_c(2S)$ and $\psi(2S)$, respectively, while the vanishing of  the matrix elements $\bra{\chi_{c{0,2}}(q)} {\bar c} \gamma^\mu c \, \ket{0}$ and $\bra{h_{c}(q,\epsilon)} {\bar c} \gamma^\mu(1-\gamma_5) c \, \ket{0}$  implies that the  $B_a \to \chi_{c0,2}\, L$ and $B_a \to h_c\, L$ amplitudes vanish in the factorization approximation.

By the factorization ansatz one has  expressions for the various decay widths.  Moreover,
for decays in  two $J=1$ mesons,  also the polarization fractions can be computed, namely  $f_L$,  the fraction of the decay width  when both the final mesons are longitudinally polarized \footnote{The two $J=1$ mesons in the final state have the same helicity since the decaying  $B_a$ is spinless.}.
The results are the following:
\begin{itemize}
\item modes where $M_{c{\bar c}}$ is either a $J^{PC}=1^{--}$   charmonium state ($J/\psi$ or $\psi(2S)$),  or a $J^{PC}=1^{++}$  P-wave $\chi_{c1}$  meson:
   \bea
\Gamma(B_a \to M_{c{\bar c}} L)&=& {G_F^2 |V_{cb} V_{cs}^*|^2 \, (a_2^{eff})^2 \, f_{M_{c{\bar c}}}^2 \over 32 \pi m_{B_a}^3} \left[F_1^{B_a \to L}(m_{M_{c{\bar c}}}^2) \right]^2 \lambda^{3/2}\left( m_{B_a}^2, m_{M_{c{\bar c}}}^2, m_L^2 \right) \,,\label{BtoS-P-width} \\
\Gamma(B_a \to M_{c{\bar c}} V)&=& {G_F^2 |V_{cb} V_{cs}^*|^2 \, (a_2^{eff})^2 \, f_{M_{c{\bar c}}}^2 \over 16 \pi m_{B_a}^3} { \lambda^{1/2}\left( m_{B_a}^2, m_{M_{c{\bar c}}}^2, m_V^2 \right) \over 8 m_V^2} \nonumber \\
&&\Big\{ (m_{B_a}+m_V)^2 [A_1^{B_a \to V}( m_{M_{c{\bar c}}}^2)]^2\,[\lambda\left( m_{B_a}^2, m_{M_{c{\bar c}}}^2, m_V^2 \right) +12 m_{M_{c{\bar c}}}^2 m_V^2]\nonumber \\
&+& { [A_2^{B_a \to V}( m_{M_{c{\bar c}}}^2)]^2 \over (m_{B_a}+m_V)^2} \lambda^2\left( m_{B_a}^2, m_{M_{c{\bar c}}}^2, m_V^2 \right) \nonumber \\
&-&2 A_1^{B_a \to V}( m_{M_{c{\bar c}}}^2) \,A_2^{B_a \to V}( m_{M_{c{\bar c}}}^2)\,(m_{B_a}^2-m_{M_{c{\bar c}}}^2- m_V^2)\,\lambda  \left( m_{B_a}^2, m_{M_{c{\bar c}}}^2, m_V^2 \right)\nonumber \\
 &+&8 m_{M_{c{\bar c}}}^2 m_V^2 {[V^{B_a \to V}( m_{M_{c{\bar c}}}^2)]^2 \over  (m_{B_a}+m_V)^2 } \lambda  \left( m_{B_a}^2, m_{M_{c{\bar c}}}^2, m_V^2 \right) \Big\} \,,\label{BtoV-width}\\
f_L(B_a \to M_{c{\bar c}} V)&=&{1 \over \Gamma(B_a \to M_{c{\bar c}} V)}
{G_F^2 |V_{cb} V_{cs}^*|^2 \, (a_2^{eff})^2 \, f_{M_{c{\bar c}}}^2 \over 16 \pi m_{B_a}^3} { \lambda^{1/2}\left( m_{B_a}^2, m_{M_{c{\bar c}}}^2, m_V^2 \right) \over 8 m_V^2} \nonumber \\
&&\Big\{ (m_{B_a}+m_V) [A_1^{B_a \to V}( m_{M_{c{\bar c}}}^2)](m_{B_a}^2-m_{M_{c{\bar c}}}^2- m_V^2) \nonumber \\&-&
\lambda\left( m_{B_a}^2, m_{M_{c{\bar c}}}^2, m_V^2 \right) { [A_2^{B_a \to V}( m_{M_{c{\bar c}}}^2)] \over (m_{B_a}+m_V)} \Big\}^2 \,;\label{BtoV-fL}
 \eea
  \item modes with a $J^{PC}=0^{-+}$  $P_{c{\bar c}}$  ($\eta_c$ or $\eta_c(2S)$) charmonium   state:
\bea
\Gamma(B_a \to P_{c{\bar c}} L)&=& {G_F^2 |V_{cb} V_{cs}^*|^2 \, (a_2^{eff})^2 \, f_{P_{c{\bar c}}}^2 \over 32 \pi m_{B_a}^3}\left[F_0^{B_a \to L}(m_{P_{c{\bar c}}}^2) \right]^2  \nonumber \\ &\times&(m_{B_a}^2-m_L^2)^2\,\lambda^{1/2}\left( m_{B_a}^2, m_{P_{c{\bar c}}}^2, m_L^2 \right)\,,\label{BtoetacS-P-width} \\
\Gamma(B_a \to P_{c{\bar c}} V)&=& {G_F^2 |V_{cb} V_{cs}^*|^2 \, (a_2^{eff})^2 \, f_{P_{c{\bar c}}}^2 \over 32 \pi m_{B_a}^3} \left[A_0^{B_a \to P}(m_{P_{c{\bar c}}}^2) \right]^2\nonumber \\ &\times&\,\lambda^{3/2}\left( m_{B_a}^2, m_{P_{c{\bar c}}}^2, m_V^2 \right) \,. \label{BtoetacV-width}
\eea
\end{itemize}
$\lambda$ is the triangular function,  $\lambda(a,b,c)=(a-b-c)^2-4bc$. Eqs. (\ref{BtoS-P-width}) and (\ref{BtoetacS-P-width})  apply to both the cases in which the light meson is  pseudoscalar or  scalar,  $L=P,S$.

Using in these expressions  the  coefficients $a_i(\mu)$  computed in renormalization group improved
perturbation theory, the experimental data are badly reproduced:   the $b \to c{\bar c}s$  induced modes  under scrutiny are colour suppressed, and the predictions of  naive factorization undershoot the data. The
most striking discrepancy is for the modes with $\chi_{c0}$ in the final state, which have a sizeable rate despite  their
amplitude vanishes in the  factorization approach.  Our strategy is to exploit the
data in Table~\ref{tab:experimental-data}  to determine an effective parameter $a_2^{eff}$
(generally  channel-dependent) and, assuming $SU(3)_F$ symmetry, to use these values to predict the flavour  related $B_s$ decays.   Since the  results depend on
the form factors,  to estimate this hadronic uncertainty  we use two sets of form factors
factors computed by variants of the QCD sum rule method \cite{Colangelo:2000dp}, the set in \cite{Colangelo:1995jv} obtained using sum rules based on the short-distance expansion, and the set in
\cite{Ball:2004ye} based on the light-cone expansion.
In the case  of  $B_s \to \phi$  and $B_s \to f_0(980)$ we use form factors determined by  light-cone sum rules  \cite{Ball:2004rg,Colangelo:2010bg}.

The  numerical inputs
$V_{cb}=0.0412\pm 0.0011$, $V_{cs}=1.04\pm0.06$,
$\tau(B^-)=(1.638\pm0.011) \,\, {\rm ps}$,  $\tau(B^0)=(1.530\pm0.009) \,\, {\rm ps} $, $ \tau(B_s)=(1.470^{+0.026}_{-0.027})\,\, {\rm ps}$, together with
the values of the meson masses, are taken from the Particle Data Group~\cite{Amsler:2008zzb}. Moreover,  from
 $J/\psi (\psi(2S)) \to e^+ e^-$ \cite{Amsler:2008zzb}  we obtain:
$f_{J/\psi}= (416.3\pm5.3)$ MeV and
$f_{\psi(2S)}=(296.1\pm2.5)$ MeV.
The decay constant of $\eta_c$ comes from $\eta_c\to 2\gamma$:
$ f_{\eta_c}=(380.0\pm87.1)$  MeV,
while only the upper bound $f_{\eta_c(2S)}<438.8$ MeV is known. In the heavy quark limit, the pseudoscalar and vector charmonia are collected in a doublet of states with degenerate masses and same decay constants.
Therefore,   $f_{\eta_c(2S)}$ can be obtained using  the relation
\begin{eqnarray}
 f_{\eta_c(2S)}&=&
 \frac{f_{\eta_c}}{f_{J/\psi}}f_{\psi(2S)}=(270.3\pm62.0)\;\;{\rm MeV};
\end{eqnarray}
symmetry breaking terms, coming from removing the meson degeneracy, are expected to cancel in the  ratio.
Since the constant $f_{\chi_{c1}}$ is not known,  for  the modes involving $\chi_{c1}$  we determine the product $a_2^{eff}  f_{\chi_{c1}}$ from  data.
$SU(3)_F$ symmetry allows to relate $B_s$ decays to those listed in Table~\ref{tab:experimental-data}:  data on   $B \to M_{c {\bar c}} K$ allow us to predict  $B_s \to M_{c {\bar c}}\eta^{(\prime)}$, while  information on $B \to M_{c {\bar c}}  K^*$ is used to predict $B_s \to M_{c {\bar c}}\phi$.  As for $B_s \to M_{c {\bar c}} f_0(980)$, they are obtained using the effective $|a_2|$ determined from $B \to M_{c {\bar c}} K$.
 The $B_s\to \eta^{(\prime)}$ form factors are related  to the analogous $B \to K$ form factors:  for a generic form factor $F$ we have $F^{B_s\to\eta}=
-\sin\theta F^{B\to K}$ and $F^{B_s\to\eta'}= \cos\theta F^{B\to K}$
where $\theta$ is the mixing angle in the flavor basis~\cite{Feldmann:1998vh}
 \begin{eqnarray}
\eta&=\eta_q \cos \theta -\eta_s \sin \theta,\nonumber\\
\eta'&=\eta_q \sin \theta +\eta_s \cos \theta,\label{eq:etamix}
\end{eqnarray}
with $\eta_q=(\bar uu+\bar dd)/\sqrt2$ and $\eta_s=\bar ss$. The
mixing angle between $\eta_q$ and $\eta_s$ can be fixed to the value measured by the KLOE Collaboration:
$\theta=\big( 41.5 \pm 0.3_{stat} \pm 0.7_{syst} \pm0.6_{th} \big )^\circ$  \cite{kloe},  which agrees with the outcome of a QCD sum rule analysis of the radiative  $\phi \to \eta^{(\prime)} \gamma$  modes \cite{DeFazio:2000my}.
%%%%%%%%%%%%%%%%%%%%%%%%%%%%%%%%%%%%%%%%%%%%%%%%%%%%%%%%%%%%%%%
\TABULAR{ccccccccccc} { \hline
mode      & $|a_2^{\rm CDSS}|$ & $|a_2^{\rm BZ}|$ & mode     & $|a_2^{\rm CDSS}|$ & $|a_2^{\rm BZ}|$  \\\hline
 $J/\psi \, \eta \, (\eta')$  & $0.40\pm0.007$  & $0.26\pm0.005$      &  $\eta_c\, \eta\, (\eta')$   &  $0.36\pm0.03$ &  $0.25\pm0.02$
 \\
 $\psi(2S)\, \eta \,(\eta')$ & $0.50\pm0.02$   & $0.31\pm0.01$     &  $\eta_c(2S)\, \eta\, (\eta')$ & $0.31\pm0.08$  & $0.21\pm0.06$
 \\
 \hline
mode       & $|a_2^{\rm CDSS}f_{\chi_{c1}}|$         & $|a_2^{\rm BZ}f_{\chi_{c1}}|$
 & mode     & $|a_2^{\rm CDSS}f_{\chi_{c1}}|$         & $|a_2^{\rm BZ}f_{\chi_{c1}}|$     \\\hline
 $\chi_{c1} \, \eta \, (\eta')$   & $0.122\pm0.006$ & $0.076\pm0.004$    &  $\chi_{c1} \, f_0$    &  $0.122\pm0.016$& $0.076\pm0.010$
 \\
  $\chi_{c1}\phi$ &  &  $0.0345\pm0.006$    \\
 \hline} { Effective Wilson coefficients $a_2^{eff}$ and combination $a_2^{eff} f_{\chi_{c1}} $ (in GeV) appearing in the decay amplitudes of the various $B_s\to M_{c\bar c}L$ modes, obtained from the $SU(3)_F$  related $B$ decay modes and using the
form factors in Ref.\cite{Colangelo:1995jv} (CDSS) and  Ref.\cite{Ball:2004ye} (BZ).\label{tab:a2eff}}
%%%%%%%%%%%%%%%%%%%%%%%%%%%%%%%%%%%%%%%%%%%%%%%%%%%%%%%%%%%%%%%
In  $a_2^{eff}$ we include the uncertainty on the form factors at $q^2=0$ and on the experimental branching ratios in Table~\ref{tab:experimental-data}. In the case of the transitions involving $\eta$ or $\eta^\prime$,  the uncertainty on the form factors is not included  since, on the basis of $SU(3)_F$,  the dependence on the form factors cancels when $B_s \to M_{c {\bar c}} \eta^{(\prime)}$ branching ratio is related to $B \to M_{c {\bar c}} K$, leaving only a dependence on the $\eta$-$\eta^\prime$ mixing angle. The resulting values of $a_2^{eff}$ are collected in Table \ref{tab:a2eff}, and the predictions for $B_s$ branching ratios  in Tables \ref{tab:Bs-BR} and \ref{tab:Bs-p-wave}. In Table \ref{tab:Bs-BR} also the available experimental data are included, with  a general  agreement with the predictions.
%
%%%%%%%%%%%%%%%%%%%%%%%%%%%%%%%%%%%%%%%%%%%%%%%%%%%%%%%%%%%%%%%
%%%%%%%%%%%%%%%%%%%%%%%%%%%%%%%%%%%%%%%%%%%%%%%%%%%%%%%%%%%%%%%
\TABULAR{cccccccc}
{ \hline
mode                &      ${\cal B}$    (CDSS)      &  ${\cal B}$ (BZ)                  & Exp.                     & mode                 & ${\cal B}$ (CDSS)              &${\cal B}$ (BZ)          \\\hline
 $J/\psi \, \eta$     & $4.3\pm 0.2$  & $4.2\pm 0.2$ &$3.32\pm1.02$  &  $\eta_c  \, \eta$  & $4.0\pm 0.7$  & $3.9\pm 0.6$ \\
 $J/\psi \, \eta'$    & $4.4\pm 0.2$  & $4.3\pm 0.2$ & $3.1\pm1.39$    &  $\eta_c  \, \eta'$ & $4.6\pm 0.8$  & $4.5\pm 0.7$ \\
  \\
 $\psi(2S)\, \eta$ & $2.9\pm 0.2$  &$3.0\pm 0.2$  &                             &$\eta_c(2S) \, \eta$  &$1.5\pm 0.8$     & $1.4\pm 0.7$ \\
 $\psi(2S) \, \eta'$& $2.4\pm 0.2$  & $2.5\pm 0.2$ &                             &$\eta_c(2S) \, \eta'$ &$1.6\pm 0.9$     & $1.5\pm 0.8$ \\
 \\
 $J/\psi\, \phi$      & ---                      &$16.7\pm 5.7$&$13\pm4$          &$\eta_c \, \phi$ &--- &  $15.0\pm 7.8$ \\
 $\psi(2S)\, \phi$ &---                       &$8.3\pm 2.7$   &$6.8\pm3.0$      &                       &     &\\
 \\
$\chi_{c1} \, \eta$ &$2.0\pm 0.2$ &$2.0\pm 0.2$ & &  $\chi_{c1}\, f_0$ &     $1.88\pm0.77$    & $0.73\pm0.30$ \\
 $\chi_{c1} \, \eta'$ &$1.9\pm 0.2$ &$1.8\pm0.2$ & &  $\chi_{c1} \, \phi$ & ---  & $3.3\pm1.3$\\
 \hline
$J/\psi \, f_0$    & $4.7\pm 1.9$   & $2.0\pm 0.8$ & $<3.26$ &  $\eta_c \, f_0$      & $4.1\pm 1.7$  & $2.0\pm 0.9$ \\
$\psi(2S) \, f_0$  & $2.3\pm 0.9$  &$0.89\pm0.36$ &    &  $\eta_c(2S) \, f_0$  & $0.58\pm 0.38$& $1.3\pm 0.8$  \\ \hline
 }
{Branching ratios ($\times 10^{4}$) of the decays $B_s\to M_{c\bar c}\, L$ using the form factors in \cite{Colangelo:1995jv} (CDSS) and
in \cite{Ball:2004ye} (BZ). The experimental results are taken from PDG \cite{Amsler:2008zzb}, except for $B_s \to J/\psi \, \eta\,(\eta')$ measured by Belle Collaboration, with the errors combined in quadrature \cite{:2009usa}.
For  $B_s\to M_{c\bar c} \, f_0$,
 the effective coefficient $a_2$ obtained from the $B\to K$ mode and the form factors  CDSS and BZ are used.   The bound  for $B_s \to J/\psi \, f_0$ is due to the Belle Collaboration \cite{talkbelle}.\label{tab:Bs-BR} }
%%%%%%%%%%%%%%%%%%%%%%%%%%%%%%%%%%%%%%%%%%%%%%%%%%%%%%%%%%%%%%%
%
Several remarks are  in order. The values of $a_2^{eff}$ derived from the form factors
in Ref.~\cite{Colangelo:1995jv} are  larger than the ones
derived from Ref.~\cite{Ball:2004ye}; they also turn out to be
channel-dependent. Their range
$(0.2-0.3)$ or $(0.3-0.5)$ is  larger than the one obtained in
the QCDF and pQCD  approaches  which undershoot  the data.
As appears from Tables
 \ref{tab:Bs-BR} and \ref{tab:Bs-p-wave},  all the modes have sizeable branching fractions, so that they are   promising
candidates for  measurements  of  $\beta_s$.
The modes involving $\eta, \,\eta^\prime, \, f_0$ present, with
respect to $B_s\to J/\psi\phi$, the advantage that
the final state is a CP eigenstate, not requiring any angular
analysis. However, the channels with  $\eta$ and $\eta'$ could be  useful only after a number of events will be accumulated, since at least
two photons are required for the reconstruction.
%%%%%%%%%%%%%%%%%%%%%%%%%%%%%%%%%%%%%%%%%%%%%%%%%%%%%%%%%%%%%%%
\TABULAR{ccccccccccc} {  \hline
mode         & ${\cal B}$      & mode         & ${\cal B}$     & mode          & ${\cal B}$\\ \hline
 $\chi_{c0}\, \eta$ & $0.85\pm0.13$ & $\chi_{c2}\, \eta$   & $<0.17$        & $h_c\, \eta$   & $<0.23$\\
 $\chi_{c0}\, \eta'$& $0.87\pm0.13$  & $\chi_{c2}\, \eta'$  & $<0.17$        & $h_c\, \eta'$  & $<0.23$\\
 $\chi_{c0}\, f_0$& $1.15\pm0.17$   & $\chi_{c2}\, f_0$  & $<0.29$        & $h_c\, f_0$  & $<0.30$\\
 $\chi_{c0}\, \phi$ & $1.59\pm0.38$ & $\chi_{c2}\, \phi$   & $<0.10(0.62\pm0.17)$ & $h_c\, \phi$   & $(<1.9)$ \\ \hline
}{Branching ratios ($\times 10^{4}$)  of $B_s$ decays into p-wave charmonia.
\label{tab:Bs-p-wave}}
%%%%%%%%%%%%%%%%%%%%%%%%%%%%%%%%%%%%%%%%%%%%%%%%%%%%%%%%%%%%%%%
%%%%%%%%%%%%%%%%%%%%%%%%%%%%%%%%%%%%%%%%%%%%%%%%%%%%%%%%%%%%%%%

As discussed  in \cite{Stone:2008ak,Colangelo:2010bg,Leitner:2010fq}, the mode $B_s\to J/\psi f_0$ has  appealing features since, compared with the $\eta$ and $\eta'$, the $f_0$ can be easily reconstructed
in the $\pi^+\pi^-$ final state, which occurs  with a large rate: ${\cal B}(f_0\to
\pi^+\pi^-)=(50^{+7}_{-8})\%$ \cite{Ablikim:2004cg}, so that  this
channel could likely  be accessed~\footnote{The quark content of $f_0$
is not completely known. Under the $\bar qq$ assignment,
this  meson might be a mixture of the isosinglet $\bar nn$ and
$\bar ss$ $(n=u,d)$ components. The mixing angle can be fixed using  experimental
information on, for instance, the decays  $J/\psi \to \phi \, f_0$ and $J/\psi \to \omega \, f_0$:
${\cal B}(J/\psi\to \phi f_0)= (3.2\pm0.9)\times 10^{-4}$,
${\cal B}(J/\psi\to \omega f_0)= (1.4\pm0.5)\times 10^{-4}$,
which might signal a nonstrange component of $f_0$ and
the consequent reduction of ${\cal B}(B_s\to J/\psi f_0)$ by
about  $30\%$}. The Belle Collaboration has provided the  measurement
\cite{talkbelle}:
\begin{eqnarray}
{\cal B}(B_s\to J/\psi f_0)\times {\cal B}(f_0\to \pi^+\pi^-)<
1.63\times 10^{-4}\,
\end{eqnarray}
which is  marginally compatible  with our prediction.

The $B_s\to \eta_c L$ are also predicted in Table~\ref{tab:Bs-BR}. Although also these channels
have sizable branching fractions, they present  the drawback of the difficult reconstruction of the $\eta_c$.

Let us now consider $B_s$ decays to  $p$-wave charmonia.
We have  stressed that, among these decays,  the only one with non vanishing amplitude in the factorization assumption is that  with $\chi_{c1}$ in the final state.
In the other cases, i.e. for modes involving $\chi_{c0,2}$ and $h_c$ collected in Table \ref{tab:Bs-p-wave},  the results are obtained
determining the decay amplitudes from the $B$ decay data by making use of the $SU(3)_F$
symmetry. In this case, the differences between the $B$ and $B_s$
decays arise from the phase space and lifetimes of the heavy mesons.
As for the mechanism inducing such processes, one possibility, put forward in \cite{Colangelo:2002mj}, is that rescattering can  be responsible of the observed branching fractions.
Among these channels,  $B_s\to \chi_{c0}\phi$ is of prime
interest and very promising for both hadron colliders and
 $B$ factories. Even though  ${\cal B}(B_s\to \chi_{c0}\phi)$ is one order of magnitude smaller
than ${\cal B}(B_s\to J/\psi\phi)$ it has appealing features,  in particular as far as the potential of the LHCb
experiment is concerned. Since the final
state consists of six charged hadrons, the particle identification
information from the RICH detectors could suppress the background.  Furthermore, the vertex detector might be particularly efficient for
these channels.

%%%%%%%%%%%%%%%%%%%%%%%%%%%%%%%%%%%%%%%%%%%%%%%%%%%%%%%%%%
\TABULAR{cccccc}
 {\hline
 Channel              &   Theory            & Experiment   \\\hline
 $J/\psi \, \phi$     & $51.3\pm5.8$  & $54.1\pm 1.7$ \\
 $\psi(2S)\, \phi$  & $41.0\pm3.7$ &              \\
 $\chi_{c1}\, \phi$ & $43.9\pm4.4$ &              \\\hline}
 {Longitudinal polarization fraction $f_L$ ($\times 10^2$) for $B_s$ decays to two $J=1$ mesons.
\label{tab:Bs-polarization}}
%%%%%%%%%%%%%%%%%%%%%%%%%%%%%%%%%%%%%%%%%%%%%%%%%%%%%%%%%%

Considering finally the  polarization fractions, in Table \ref{tab:Bs-polarization} we collect our predictions for the longitudinal polarization fractions $f_L$ for the modes with two $J=1$ mesons in the final state. There is agreement with experiment  for $B_s \to J/\psi \phi$, the only mode for which data on $f_L$ are available. This is at odds with the case of a few  suppressed $B$ decays to two light vector mesons,  in which the  experimental datum is not reproduced assuming factorization.
Actually, it should be noticed that naive and generalized factorization provide  the same result in the case of the polarization fractions since they differ only for the value of $a_2^{eff}$ which cancels in the ratio defining a polarization fraction. In order to modify the prediction for $f_L$ one should either consider approaches in which the three polarization fractions (the longitudinal and the two transverse ones) are weighted by different Wilson coefficients or invoke again other mechanisms such as rescattering. The first case is realized in QCD factorization and in pQCD. As for rescattering, it has been proposed as a solution to the puzzle of the polarization fractions in the case of $B$ decays to two light vector mesons, when the considered process is suppressed as in the case of the penguin induced mode $B \to \phi K^*$ \cite{Colangelo:2004rd}.

\section{New physics in nonleptonic $B_s$ decays:  general analysis}
\label{sect:np}

As  mentioned in the Introduction, hints of deviations from SM  predictions have recently been found in $B_s$ phenomenology, hence it is worth
 considering  the effects of new physics in the $B_s $ sector, which  may show up   in mixing and/or in decay amplitudes.

New physics in $B_s -{\bar B}_s$ mixing can modify  the mixing phase $\beta_s$.   We refer to this phase as to $\beta_s^{eff}$, which contains SM as well as NP contributions: $\beta_s^{eff}=\beta_s^{SM}+\beta_s^{NP}$. This effect is the same for all decay modes, and simply shifts the value of $\beta_s$.
On the other hand, NP in the decay amplitudes can affect  various channels in  different ways,  even for modes induced by the same quark transition, as we specify in the following.

Let us discuss the possibility that experimental results for nonleptonic $B_s$ decays deviate from the predictions given in the previous Section. Such predictions rely on $SU(3)$ flavour symmetry and on experimental data on  corresponding $B_d$ decays, in which no NP effects have been detected at the present level of accuracy. Deviations in $B_s$ decay rates with respect to the predictions  could be  due to a  violation of $SU(3)_F$ symmetry, which is generally  expected at a few percent level.
 However, there is the more exciting possibility of  deviations  due to
 NP effects  with small contributions  in  $B_d$ oscillations and decays  and detectable  contributions in $B_s$,  an eventuality which is  interesting to consider  for the
modes studied in this paper.  Such modes receive contribution both from tree level and  loop diagrams,  so that one would  expect NP to affect them negligibly. However, there are scenarios in which the contribution of new particles in loop diagrams can be competitive with the SM tree level diagrams. This is the case, for example,  of  supersymmetric  scenarios in which one loop gluino exchanges for $b \to s$ transition  could give a sizeable contribution to the $b \to c {\bar c}s$ induced modes. This would affect the branching ratios of such modes, and  the CP asymmetries, since new phases  could arise through the soft supersymmetry breaking terms. To avoid constraints on such phases  from existing limits on dipole electric moments, one should consider flavour dependent phases.

Here we do not focus on a specific NP model, rather we parameterize the effects of new physics in a general way, i.e. in terms of an amplitude, a weak and a strong phase,  and discuss how these quantities can be constrained by experimental data on the modes considered  above.

In a  customary notation, ${\cal A}_f$ is the amplitude  for $B_s \to f$  decay to a generic final state $f$ (CP eigenstate, common to $B_s$ and ${\bar B}_s$) which, in our case, is of the kind $M_{c{\bar c}} L$ \footnote{In the case of two vectors in the final state $f$ denotes one of the final state components being CP eigenstate.}.
The corresponding ${\bar B}_s$ decay amplitude is denoted as ${\bar{\cal A}_f}$.
Being interested in CP asymmetries, we introduce the  quantity
\be
\lambda_f=e^{-2i\beta_s^{eff}}\left({ {\bar {\cal A}_f} \over {\cal A}_f} \right) \,\,\, ,
\label{lambdaf} \ee
\noindent in terms of which one can write the { mixing induced} CP asymmetry $S_f$ and the { direct} CP asymmetry $C_f$:
\be
S_f={2 \Im (\lambda_f) \over 1+|\lambda_f|^2} \,\,\, , \hspace {1cm} C_f={1-|\lambda_f|^2 \over 1+|\lambda_f|^2} \,\,. \label{sf-cf}
\ee
Assuming that there is a single dominant NP amplitude (or that all NP amplitudes have the same weak and strong phases relative to the SM), we write:
\bea
{\cal A}_f&=& |{\cal A}_f^{SM}|+ |{\cal A}_f^{NP}|\,e^{i \theta_{NP}}\,e^{i \delta_{NP}} \nn \\
&=&  |{\cal A}_f^{SM}|(1+ R\,e^{i \theta_{NP}}\,e^{i \delta_{NP}}) \,\,\, ,\label{af} \eea
where $R=\displaystyle{|{\cal A}_f^{NP}| \over |{\cal A}_f^{SM}|}$ is the ratio of the modulus of the NP amplitude and that of the SM one, while $\theta_{NP}(\delta_{NP})$ is the strong (weak) NP phase with respect to the SM part.
Our working hypothesis is that no NP affects $B$ decays:  actually,  the  new weak phase $\delta_{NP}$ would be the same in $B$ and $B_s$ decays, depending only on the underlying quark transition, however $R$ and $\theta_{NP}$ depend on the matrix elements of the operators between initial and final states, which can be different.

Using the definition in eq. (\ref{af}) and considering that: $\bar {\cal A}_f= |{\cal A}_f^{SM}|(1+ R\,e^{i \theta_{NP}}\,e^{-i \delta_{NP}}) $, we  get for the CP-averaged branching fraction
\be
{\cal B}^{exp}={\cal B}^{SM}\left[ 1+2R\cos(\theta_{NP})\cos(\delta_{NP}) +R^2 \right] \,\,\, , \label{br-np} \ee
where from now on we shall omit the label $f$  in  the quantities in eq. (\ref{br-np}) though they are  channel dependent (except for $\delta_{NP}$ which is the same for all the modes induced by the same underlying quark transition).
From eq. (\ref{lambdaf}), we obtain:
\be
\lambda_f=e^{-2i\beta_s^{eff}} {1+R e^{i(\theta_{NP}-\delta_{NP})} \over 1+R e^{i(\theta_{NP}+\delta_{NP})} } \label{lf-ris} \ee
so that\bea
S_f &=& -\eta_f {\sin(2 \beta_s^{eff})+2R\cos \theta_{NP} \sin(2 \beta_s^{eff}+\delta_{NP}) + R^2 \sin(2 \beta_s^{eff}+2 \delta_{NP}) \over 1+2R \cos \theta_{NP}  \cos \delta_{NP} +R^2 } \,\,\,, \label{sf-ris} \\
C_f&=& {2R \sin \theta_{NP}  \sin \delta_{NP}
\over 1+2R \cos \theta_{NP}  \cos \delta_{NP} +R^2 } \label{cf-ris} \,\,, \eea
$\eta_f$ being the CP eigenvalue of the final state $f$.
In absence of NP we  recover the SM results $\lambda_f=e^{-2i\beta_s}$, $S_f = -\eta_f
\sin(2 \beta_s)$, $C_f=0$. These results would also hold if NP contributes only  to $B_s-{\bar B}_s$ mixing with $\beta_s \to \beta_s^{eff}$.

The three equations (\ref{br-np}), (\ref{sf-ris}), (\ref{cf-ris}) allow  to determine the  NP parameters $R$, $\theta_{NP}$, $\delta_{NP}$, once experimental data on ${\cal B}(B_s \to f)$, $S_f$ and $C_f$ are available for a given final state $f$.
Assuming  $R \ll 1$,  we obtain:
\bea
\theta_{NP}&=&ArcTan \left( C_f \over {\tilde S}_f \right) \,\,\, , \label{thetaNP} \\
\delta_{NP}&=& ArcTan \left[{(1+\Sigma) \over \Sigma} {\tilde S}_f \right] \,\,\, , \label{deltaNP} \\
R&=& {\Sigma \over 2\cos(\theta_{NP})\cos(\delta_{NP})} \,\,,\label{R-ris} \eea
where $\Sigma$ and ${\tilde S}_f$ parametrize deviations from the SM:
\be
\Sigma= {{\cal B}^{exp} \over {\cal B}^{SM}}-1\,\,\, ,  \label{sigma-def}
\ee
\be
{\tilde S}_f= { -\eta_f S_f -\sin(2 \beta_s^{eff} )
\over \cos (2 \beta_s^{eff})} \label{sftilde-def} \,\,.
\ee

\begin{figure}[htth]
\begin{center}
\includegraphics[scale=0.65]{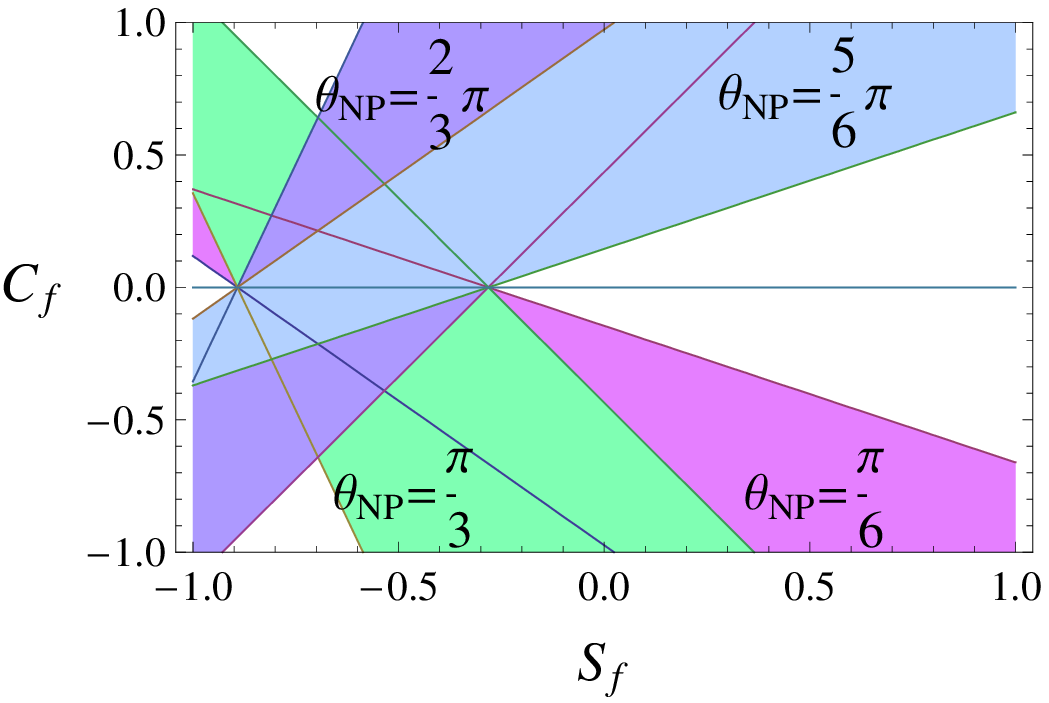}\hspace{0.6cm}
\includegraphics[scale=0.65]{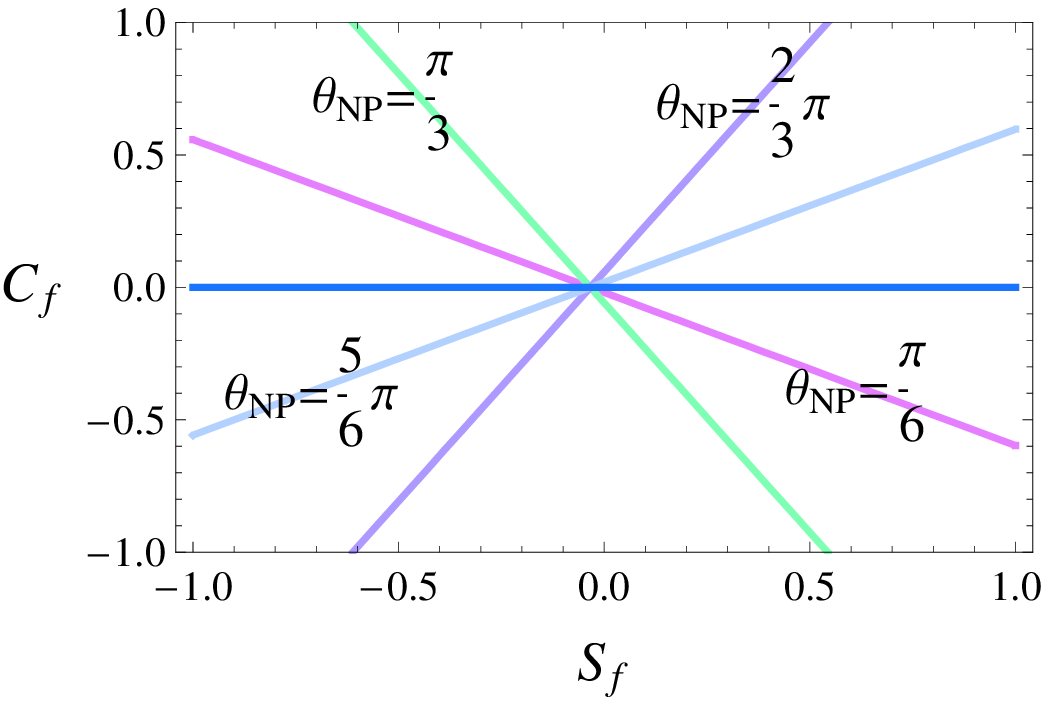}
\caption{Direct CP asymmetry $C_f$ versus the mixing-induced CP asymmetry $S_f$ for several values of the strong phase $\theta_{NP}$.}\label{CfvsSf}
\end{center}
\end{figure}

In  fig. \ref{CfvsSf} we plot the direct CP asymmetry $C_f$ versus the mixing-induced CP asymmetry for several values of the strong phase $\theta_{NP}$, using the relation (\ref{thetaNP}). Once for a given channel $f$ there will be data available for ($S_f$, $C_f$) one could find a range for $\theta_{NP}$.
The two panels in fig. \ref{CfvsSf} are obtained assuming different values for the $B_s-{\bar B}_s$ mixing phase, which in left panel is fixed to $2\beta_s^{eff}=0.77 \pm 0.37$ {\rm rad} (one of the values obtained by HFAG averaging the experimental results provided by the Tevatron Collaborations CDF and D0) while in the right panel
is fixed  to the SM value $\beta_s=0.017$ {\rm rad} (since no errors are attached to the SM value,  the various regions shrink to lines).
In these figures only positive values of $\theta_{NP}$ have been considered, since $C_f(-\theta)=C_f \left(\displaystyle{\pi \over 2}+\theta \right)=-C_f(\theta)$.
%%%%%%%%%%%%%%%%%%%%%%%%%%%%%%%%%%%%%%%%%%%%%%%%%%%%%
%%%%%%%%%%%%%%%%%%%%%%%%%%%%%%%%%%%%%%%%%%%%%%%%%%%%%

\begin{figure}[htth]
\begin{center}
\includegraphics[scale=0.65]{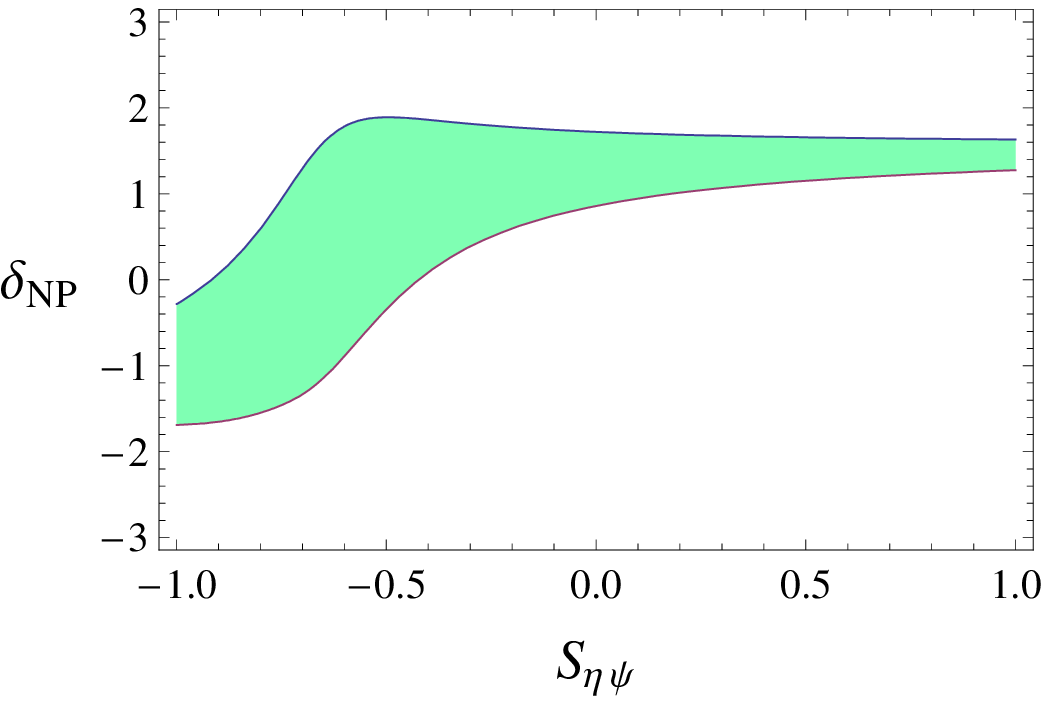}\hspace{0.6cm}
\includegraphics[scale=0.65]{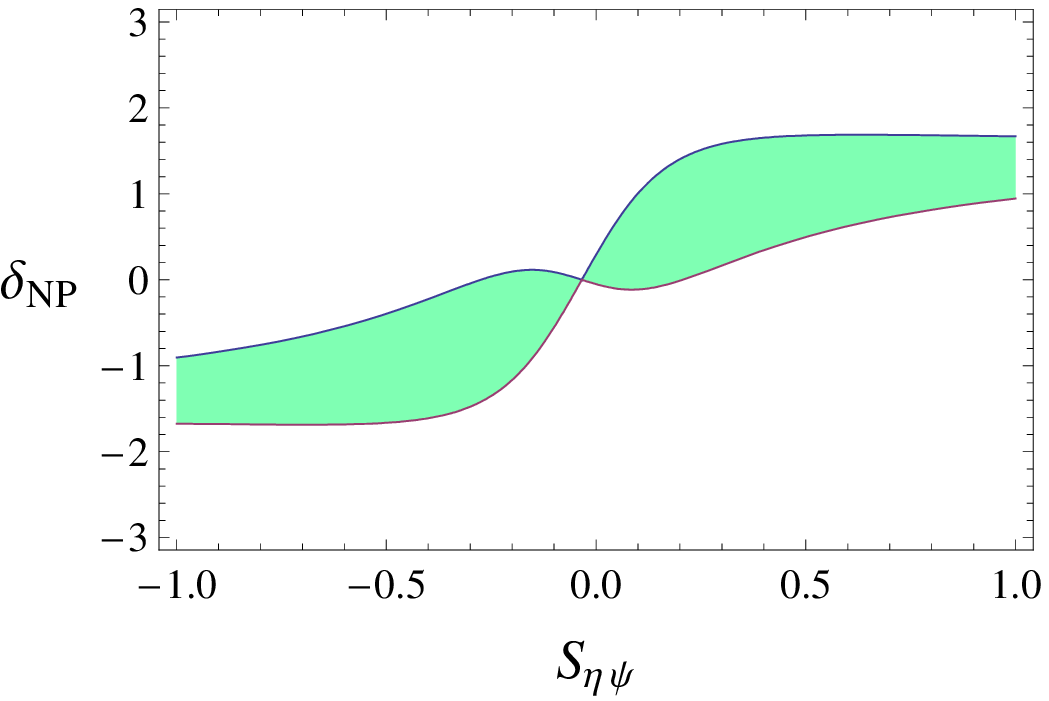}
\caption{Weak phase $\delta_{NP}$ versus $S_{J/\psi \eta}$ in the  case $\theta_{NP}=0$ and for $\Sigma$ obtained from the values of ${\cal B}^{exp}$  (left) and
${\cal B}^{SM}$ (right)  in Table 3.}\label{deltaplot}
\end{center}
\end{figure}

Eq. (\ref{deltaNP}) allows to determine the weak phase $\delta_{NP}$ using the measured ${\cal B}(B_s \to f)$ and  $S_f$. To appreciate how this could be obtained,
we consider  the final state $J/\psi \eta$,  with $\eta_f=1$,  and  plot in Fig.\ref{deltaplot}  $\delta_{NP}$ versus $S_{J/\psi \eta}$ for $\Sigma$  corresponding to the values  ${\cal B}^{exp}=3.32 \pm 1.02$ (see Table \ref{tab:Bs-BR}) and
${\cal B}^{SM}=4.25 \pm 0.28$ (the average of the (CDSS) and (BZ) predictions   in Table \ref{tab:Bs-BR}).
%%%%%%%%%%%%%%%%%%%%%%%%%%%%%%%%%%%%%%%%%%%%%%%%%%%%%
%%%%%%%%%%%%%%%%%%%%%%%%%%%%%%%%%%%%%%%%%%%%%%%%%%%%%

Last, we  consider eq. (\ref{R-ris}).  In this case,  all the three observables ${\cal B}(B_s \to f)$,  $S_f$ and $C_f$ are required to constrain $R$.

As a final remark,  once experimental information is  available about  the three observables ${\cal B}(B_s \to f)$,  $S_f$ and $C_f$ in at least two decay modes , we could constrain also $\beta_s$, since only the NP parameters $R$ and $\theta_{NP}$ are channel-specific, while $\delta_{NP}$ is the same for all the modes induced by the weak transition $b \to c {\bar c} s$. Hence, measuring in  two channels the six  observables $({\cal B}(B_s \to f_1),\,S_{f_1},\,C_{f_1})$ and $({\cal B}(B_s \to f_2),\,S_{f_2},\,C_{f_2})$ it would be possible to determine  $R_1,\,\theta_{NP,1}$, $R_2,\,\theta_{NP,2}$, $\delta_{NP}$ and  $\beta_s$.

\section{Conclusions}
Recent results in the $B_s$ sector  require  efforts to identify  the most promising ways to unveal new physics. We have considered decay channels induced by the $b \to c {\bar c}s$ transition, using generalized factorization together with $SU(3)_F$ symmetry to predict their branching fractions in the Standard Model.
Modes with a charmonium state plus $\eta$, $\eta^\prime$, $f_0(980)$ are  interesting, since they are  CP eigenstates and do not require angular analyses. In particular, the case of $f_0$ is particularly suitable in view of  its easier reconstruction in the  $\pi^+ \pi^-$ mode.

If NP affects the $B_s$ sector, it can either contribute to the $\Delta B=2$ induced mixing amplitude, in a channel independent way, modifying the value of the mixing phase with respect to its SM value, or modify the decay amplitudes, in a way that can vary from one channel to the other. In this case, the mixing induced CP asymmetry $S_f$ would no more be equal to $-\eta_f \,\sin 2 \beta_s$ and the direct CP asymmetry $C_f$ would differ from zero, conditions which instead hold in the SM or in the case that NP affects only the oscillation process. Both cases can also happen.
We have considered a general parametrization of new physics in which the second condition holds, fixing $\beta_s$ either to the average of CDF and D0 results or to its SM value. We have shown that,  with data  on the branching ratio and the two CP asymmetries $S_f$ and $C_f$ for a given final state $f$, it will be possible to constrain NP parameters.

\acknowledgments
FDF thanks A.J. Buras for very useful discussions.
 This work was supported in part by the EU contract No. MRTN-CT-2006-035482, "FLAVIAnet".

%%%%%%%%%%%%%%%%%%%%%%%%%%%%%%%%%%%%%%%%%%%%%%%%%%%%%%

\end{document}